\documentclass[letterpaper,dvipsnames,twocolumn,10pt]{article}
\usepackage{usenix}

\usepackage{nopageno}
\usepackage{tikz}
\usepackage{cite}[sort]
\usepackage{amsmath}
\usepackage[labelfont=bf]{caption}
\usepackage{pgfplots}
\usepackage{pgfplotstable}

\usepackage[compact]{titlesec}
\usepackage{booktabs}

\usepackage{filecontents}

\usepackage{enumerate}

\usepackage{subcaption}
\usepackage{float}
\usepackage{cleveref}
\usepackage{amssymb}
\usepackage{caption}
\usepackage[scaled=0.85]{beramono}
\usepackage{listings}
\usepackage{hyperref}
\usepackage{xcolor}

\newfloat{lstfloat}{htbp}{lop}
\floatname{lstfloat}{Listing}

\newcommand\code[1]{\lstinline$#1$}

\lstdefinelanguage{Pseudocode} {
 keywords=[1]{class,def,import,if,else},
 keywordstyle=[1]\color{BurntOrange}\bfseries,
 keywords=[2]{ModuleAPI},
 keywordstyle=[2]\color{BrickRed}\bfseries,
 keywords=[3]{register,push,read,read\_event,read\_last\_n,network},
 keywordstyle=[3]\color{NavyBlue}\bfseries,
 basicstyle=\small\ttfamily,
 sensitive=false,
 identifierstyle=\color{black},
 comment=[l]{\#},
 commentstyle=\color{gray}\ttfamily\bfseries,
 showstringspaces=false,
 string=[s]{"}{"},
 stringstyle=\color{BrickRed}\ttfamily,
}

\newcommand\MZ[1]{}

\begin{document}

\date{}

\title{\Large \bf
  Extricating IoT Devices from Vendor Infrastructure with Karl}

\author{
{\rm Gina Yuan, David Mazières, Matei Zaharia}\\
Stanford University
} 

\maketitle

\begin{abstract}

Most consumer IoT devices are vertically integrated with cloud-side
infrastructure.  Such architectures present enormous risk to user data,
exacerbated by vendor heterogeneity and the inability for users to
audit cloud-side activity.
A more promising approach would be to leverage local hardware, providing
users control over how their data is processed and why it can be
shared with other devices or the Internet.


Karl is a new smart-home framework designed to host IoT computation
and storage on user-chosen devices.
A key insight in Karl's modular programming model is that a familiar
interface (inspired by serverless) can capture most
modern cloud-side IoT components under a single framework,
which executes modules agnostic of hardware location.
While local hosting eliminates many flows, modularity enables all
remaining flows to be justified using fine-grained primitives.
We introduce two IoT security mechanisms: \emph{pipeline permissions} that
permit device data to be shared given some justification and
\emph{exit policies} that block flows unless specific conditions are met.
We evaluate Karl through two end-to-end applications.

\end{abstract}

\section{Introduction}

This paper presents Karl, a new smart home framework that eliminates
dependence on vendor services by making local hosting practical.  The
tight integration between vendor services, companion apps, and modern
IoT devices has increased the size of the smart home attack
surface~\cite{alrawi2019sok}.
For example, most smart speakers parse audio commands such as
``turn on the light'' using a cloud service, even when they
pertain only to local devices; doorbell cameras send video to the cloud;
and even smart locks work through cloud services.
Many devices are hard to update or lose vendor support over time,
leaving them with software vulnerabilities.
The need for devices to interoperate through
``hubs'' like HomeKit~\cite{dixon2012operating,smartthings,homekit}
can enable cross-device attacks.
Overall, the dependency on vendor services has led to a large range of attacks on smart home devices~\cite{ubiquiti2021krebs,roberts2021parler,bitdefender2021ring,sears2019fox,greenberg2018echo,larson2017fda,garlati2016owlet,antonakakis2017understanding,zdnet2021osborne}.

\begin{figure}[t]
  \centering
  \begin{subfigure}[b]{0.85\linewidth}
    \centering
    \includegraphics[width=\columnwidth]{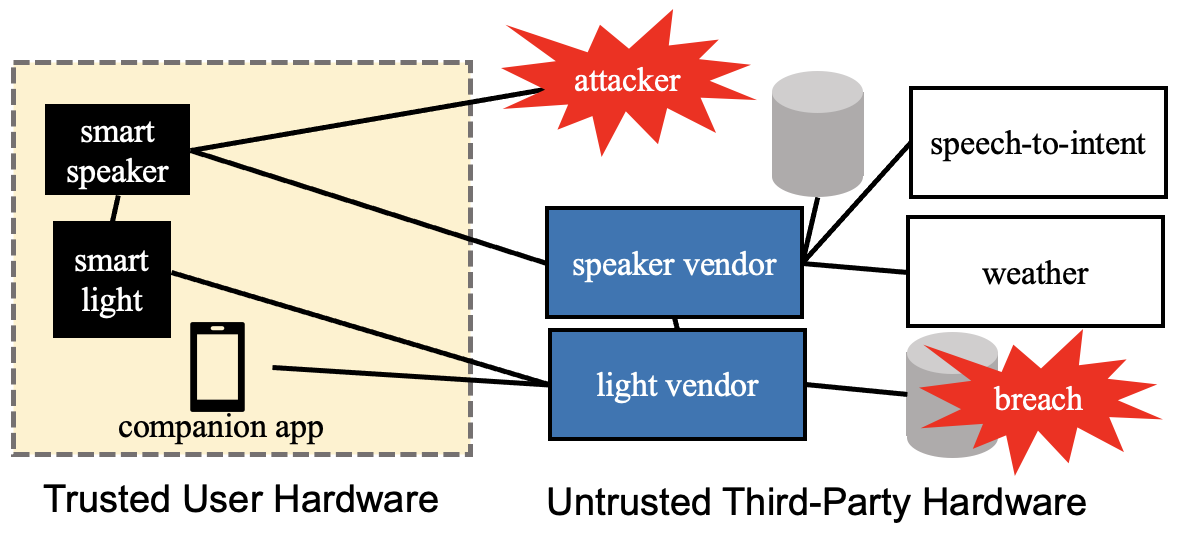}
    \caption{The modern IoT ecosystem relies heavily on remote hardware.}
    \label{fig:overview-ss}
  \end{subfigure}

  \begin{subfigure}[b]{0.85\linewidth}
    \centering
    \includegraphics[width=\columnwidth]{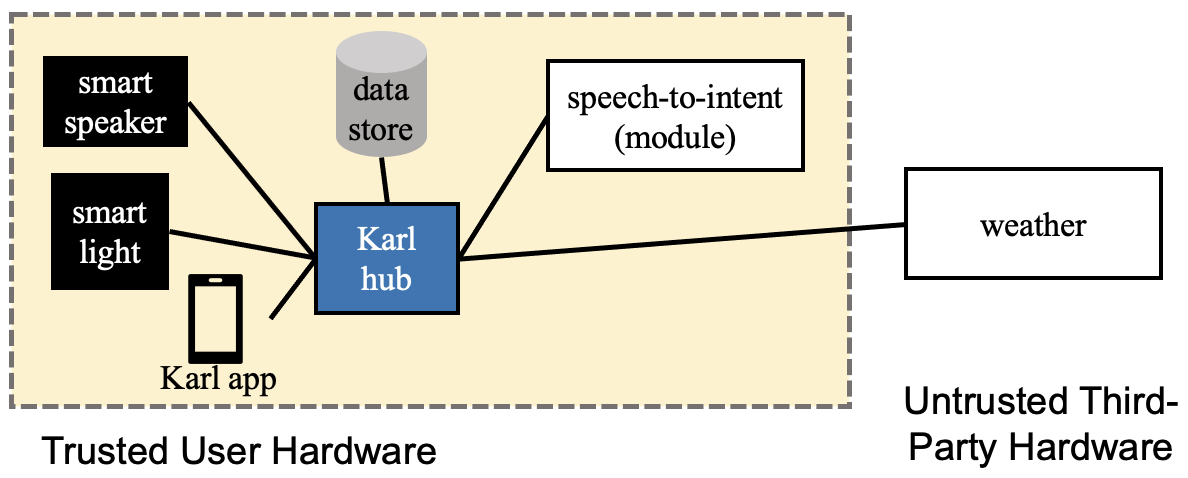}
    \caption{Karl executes most functionality on local hardware.}
    \label{fig:overview-karl}
  \end{subfigure}
  \caption{Division of device functionality on local and remote hardware. In this example, the smart speaker handles two speech commands: one that asks the weather and one that turns on a light.}
  \label{fig:overview}
\end{figure}

Theoretically, an architecture in which user data never leaves user
hardware offers better privacy and security.
For many functions, such as turning on the light by voice,
there is no need to send data remotely.
Indeed, enterprise-grade IoT devices often have purely local control.
Unfortunately, the consumer setting creates several challenges to
local hosting:  Devices are typically inexpensive and computationally
constrained.  Home networks
prevent incoming connections, making it hard to connect companion apps
directly to a device such as a camera.
Users are less sophisticated, unable to
configure servers or firewalls, and have difficulty understanding the
security implications of configuration choices.  And of course the
lower price point of consumer devices makes almost any level of
individual customer support untenable.  The result is an over-reliance
on vendor services, to the point that even locally hosted controllers
such as Home Assistant~\cite{homeassistant} have a depressingly high fraction of
integrations going through the cloud to communicate with local devices.



\begin{figure}[t]
  \centering
  \begin{subfigure}[b]{0.38\linewidth}
    \centering
    \includegraphics[width=0.9\columnwidth]{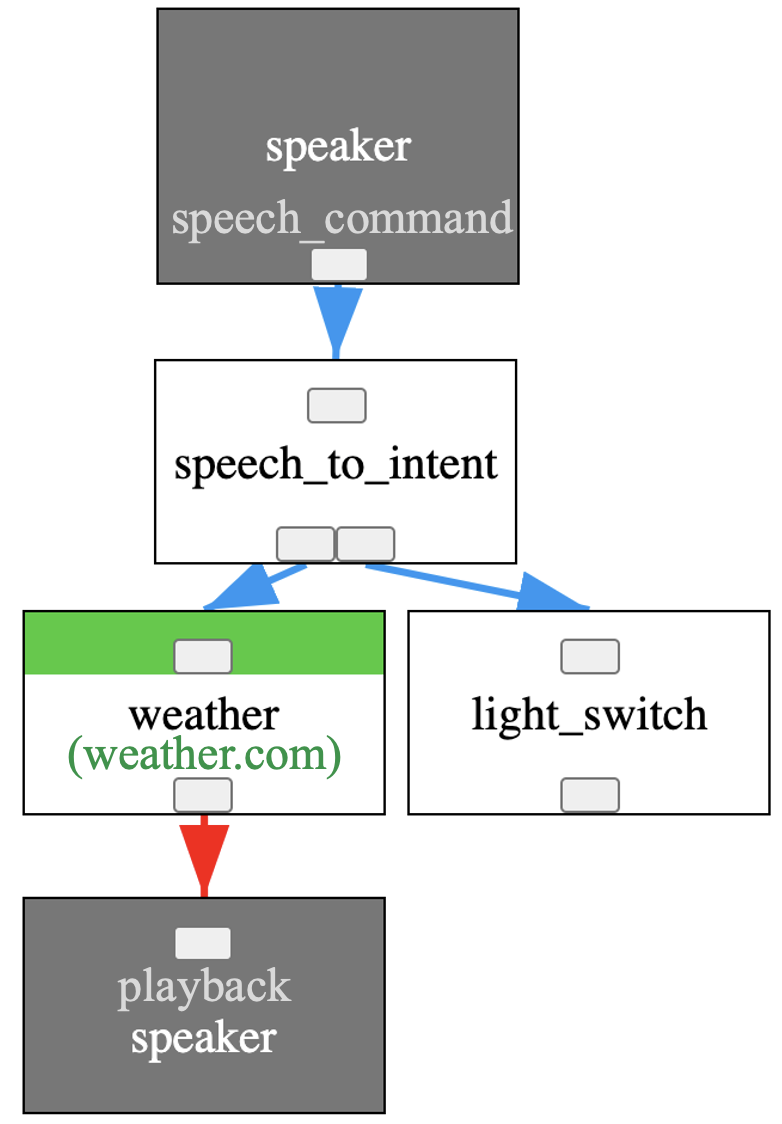}
    \caption{Example dataflow graph of a smart speaker.}
    \label{fig:dataflow-graph}
  \end{subfigure}
  \begin{subfigure}[b]{0.58\linewidth}
    \textbf{Pipeline Permission}\\
    \small{
    \texttt{
    \textcolor{BurntOrange}{speaker.speech\_command}
    $\rightarrow$ speech\_to\_intent
    $\rightarrow$ weather
    $\rightarrow$ \textcolor{ForestGreen}{weather.com}}}\\

    \textbf{Exit Policy}\\
    \small{
    Tag: \texttt{\textcolor{BurntOrange}{speaker.speech\_command}}\\
    Exit Policy: \texttt{speech\_to\_intent}\\
    }
    \caption{A generated pipeline permission and a user-specified exit policy.}
    \label{fig:security-mechanisms}
  \end{subfigure}
  \caption{
  The pipeline permission shows that the speaker wishes to send all outgoing
  data through the \texttt{speech\_to\_intent} module, which
  converts speech to text intents; then weather-related intents
  go to \texttt{\textcolor{ForestGreen}{weather.com}}. Separately,
  the user could add an exit policy saying ``all speech commands need to go
  through \texttt{speech\_to\_intent}''.}
\end{figure}

\begin{table*}[ht]
\center
\small
  \begin{tabular}{lll}
    \toprule
    Category&Justification&Smart Speaker Example\\
    \midrule
    Offload computation & weak local hardware & ML e.g., speech to intent\\
    Offload storage & scalability, fault tolerance & record notable audio events\\
    Remote access & minimize system administration of NAT and firewalls & listen to remote audio feed\\
    Pull data & data not available at production, Internet-scale knowledge &weather queries, firmware updates\\
    Push data & aggregate analytics on data & bug reports, training data, analytics\\
  \bottomrule
\end{tabular}
  \caption{Classifying device network accesses and their justifications.}
  \label{tab:class}
\end{table*}

Fig.~\ref{fig:overview} shows Karl's architecture.
Karl leverages local hardware to move processing and storage out of
devices and centralized vendor services and onto a user-controlled
device such as an old PC.  It offers a simple but
expressive programming model based on \textit{modules} that
process data, such as speech-to-intent, and communicate with each
other through a key-value \textit{data store}. 
Each module executes in isolation, similarly to a
serverless function~\cite{amazon-lambda}, and each data
and network access is validated by the locally-hosted \textit{Karl hub}.
The hub also serves \textit{Karl apps}
directly to the user's phone.

While existing IoT hubs enable access control between different devices
and the Internet, this has proven too
coarse-grained~\cite{celik2018sensitive,fernandes2016security,lee2017fact,dixon2012operating}.
For example, SmartApps are always granted full access to a device
even when they only require limited access~\cite{fernandes2016security}.
Karl, by contrast, enforces security at the module granularity.  It
represents the smart home as a dataflow graph, where device data flows
through modules to other devices and the network
(Fig.~\ref{fig:dataflow-graph}).  Users interact with the graph
through a UI that visually
represents the policy.  Most importantly, whereas users cannot audit
cloud services or enforce their terms of
service~\cite{jensen2004privacy,earp2005examining}, Karl can enforce
the policies it presents to users on all locally hosted code.

While local hosting eliminates many data flows, modularity enables all
remaining flows to be \textit{justified} using fine-grained
primitives.  Karl provides two security mechanisms
(Fig.~\ref{fig:security-mechanisms}):  First, Karl automatically
generates easily-reviewable \emph{pipeline permissions} that describe
\textit{why} device data can flow to another device or to the network
(e.g., the speaker must share a location to query the weather).
Second, Karl allows users to set \emph{exit policies} on data to block
flows unless specific conditions based on modules are met
(e.g., data derived from raw audio can only exit if it
has first been transformed to a textual intent).
Compared to existing IoT frameworks, Karl can provide more ubiquitous and 
meaningful privacy guarantees.


Karl tackles four major challenges to achieve its goals of eliminating
dependence on vendor services and enabling expressive privacy policies
on all remaining network accesses:

\paragraph{Supporting Complex Device and Service Interactions}
The first challenge is accounting for the entire attack surface of
modern devices (Table~\ref{tab:class}).
Consider a smart speaker such as the Amazon Alexa or Google Home.
Due to hardware constraints, the speaker must offload some computation
and storage.
When the user interacts with the speaker through a companion app,
they must go through a cloud endpoint mediated by the vendor.
The speaker may also download firmware updates, share analytics,
or otherwise pull and push data to the Internet.
Each network interaction adds a security risk.

Existing frameworks were not designed for the network security demands of
modern IoT.
Industry hubs like SmartThings~\cite{smartthings}
and research frameworks built on top
of it~\cite{tian2017smartauth,jia2017contexiot}
still require vendors to host their own services.
For example, Alexa's SmartThings integration serves only to
connect to Amazon servers.
Frameworks like FlowFence~\cite{fernandes2016flowfence} mediate
remote connections through the hub,
but still permit flows to leak data when not fundamentally required.
For example, FlowFence explicitly allows sharing a door lock's state with the
Internet since it is required for the mobile companion app.

In comparison, Karl moves modern functionality
to trusted hardware under a simple but expressive
programming model.
Karl executes its modules locally and also hosts the data store
on local hardware.
The Karl hub also runs and serves companion apps directly to a user's phone
through a user's router.
Flows that use an external service, such as downloading firmware updates,
can still contact those services but \textit{only if the user allows them}
with Karl policies.
These basic abstractions enable most device-to-device and companion app
interactions to run on Karl-controlled hardware.

\paragraph{Usable Privacy Policies}
The second challenge is expressing privacy policies
when a device \textit{must} access the Internet.
Existing techniques are either obfuscated in cloud infrastructure or based on
coarse-grained security primitives such as device-to-device access control.
In comparison, local hosting enables Karl to enforce ubiquitous policies that
consider all components of modern IoT functionality, as opposed to blindly
trusting multiple vendors.
Modules also enable finer-grained privacy policies with semantic meaning
such as requiring audio data to be converted to text.
Thus Karl's programming model empowers users to express and enforce
policies that correspond to more intuitive privacy guarantees.

We additionally co-design functionality and policies in Karl
such that modifying one reflects in the other. In particular, Karl
enforces user privacy policies by modifying the dataflow graph that
represents device functionality to comply with pipeline permissions
and exit policies, such as by disabling network access or deleting an edge.
Karl treats the generated graph as a
declarative policy, automatically labeling data with non-hierarchical
tags in the data store and enforcing mandatory access control on the
edges.  This mechanism combines the power of information
flow with the simplicity of access control to enforce useful privacy policies
at a fine granularity.

\paragraph{Performance}
A third challenge is maintaining performance for
real-time interactions and compute-heavy tasks without
amortizing costs on cloud hardware.
By using local hardware, Karl can minimize
latencies for many interactions~\cite{satyanarayanan2017emergence} and
even outperform cloud-based services.
For heavy compute tasks such as ML, which typically run as
long-lived services in the cloud, Karl accounts for the
high initialization times of its stateless modules using a variant of
speculative execution.
Overall, we find that Karl has acceptable performance.

\paragraph{System Administration}
Though Karl is best hosted on local hardware
for privacy, end users may prefer the durability and
scalability of the cloud.
We show it is feasible to deploy Karl in the cloud but
on \textit{user-managed hardware},
a similar approach to DIY Hosting~\cite{palkar2017diy}.
The cloud deployment increases network latency and costs money,
but self-hosting Karl
in the cloud (e.g., using Amazon Web Services) is still
comparable in cost to popular IoT cloud subscriptions,
and improves privacy because the user need only trust their own
cloud provider rather than vendors and their
providers.\\



We implement and evaluate a functioning prototype of Karl.
We port 4 home IoT devices---a security
camera, smart speaker, light bulb, and occupancy sensor.
We implement several applications in
which these devices interact with each other and third parties
outside the home network.
We analyze several privacy properties that users may
desire but cannot be enforced in existing frameworks,
and demonstrate how they are intuitive to express in Karl.
In addition, we demonstrate that Karl offers
reasonable performance and can improve the latency
of some interactive applications.
Finally, we analyze the tradeoffs of self-hosting Karl in the cloud.

In summary, our contributions are:
\begin{itemize}

  \item We introduce Karl's modular programming model,
    which makes it practical to support the rich interactions of modern
    devices on local, user-managed hardware.

  \item We introduce \textit{pipeline permissions},
    which restrict devices to justifiable data flows based on
    finer-grained primitives with semantic meaning.

  \item We introduce \emph{exit policies} as a way to capture
    end-to-end confidentiality policies users care about and avoid
    unintended consequences of pipeline permissions.

  \item We present and evaluate an implementation of Karl on
    applications spanning 4 devices and 10 modules.

\end{itemize}

\section{Security Properties}

The stakeholders in the Karl ecosystem are hardware vendors, module
developers, and end users.  We assume end users properly install Karl,
configure secure login credentials (through the Karl hub UI),
and that Karl can mediate all Internet traffic.  Some
hardware vendors and module developers may be compromised or
negligent, however, leading to vulnerable components.  Attackers
attempt to leverage these vulnerable components---possibly in
collusion with their compromised creators---so as to violate end-user
privacy or actuate smart-home controls in undesired ways.

Karl's security guarantee is that it restricts inter-component
communication to the dataflow graph specified by the user's pipeline
permissions and exit policies.  The benefit of this model is that the
presence of a few secure modules can significantly restrict the
consequences of compromised components.  For instance, the only path
through which training data can leave the network may go through a
``statistics'' module that restricts communication to
a trusted vendor's analytics endpoint.
In this case, a camera with malicious firmware cannot leak
video to outside parties on the wide-area network.

Some vendors may host cloud services, but this will be apparent in the
pipeline permission.  In many cases, the communication can be mediated
by simple, well-known, and trustworthy modules that provide meaningful
security guarantees.  For example, firmware updates should be mediated
by a data pump module that ensures devices receiving firmware cannot
send large amounts of locally recorded data back upstream.

Pipeline policies enable visual verification that devices gain
approximately least privilege, simplifying the process of
determining that the permissions suit the functionality.  In practice,
only a small fraction of users may actually object to
unwarranted permissions, but the hope is that this is enough to flag
dangerous devices for the rest of the ecosystem.

Karl does not prevent covert channels that leak information through
resource utilization.  Nor does it prevent side-channel attacks, such
as cache timing, microarchitectural data leaks, or network timing
attacks.  However, such attacks are typically
difficult to disguise in plausibly legitimate source code.  Hence, we
assume most hardware vendors and module developers are
well-intentioned with reputations to uphold, and would not risk
wall-banging attacks in production code.

\section{Motivating Examples}
\label{sec:overview}

To motivate the need for a modern smart home framework with strong privacy
guarantees, we start with a comparison of popular IoT frameworks in
industry (Samsung SmartThings), the open-source community (Home Assistant),
and research (FlowFence).
We base our comparison off the smart speaker example from the introduction
and discuss several privacy properties a user might desire.
For each framework, we analyze how a developer would implement the
smart speaker, and how the user would enforce the
following privacy properties:

\begin{enumerate}
	\setlength{\itemsep}{-1pt}
	\item The speaker can share data with \texttt{weather.com} if
	I ask about the weather, but not the raw audio.
	\item The speaker can share data with the light if I ask it to
	turn the light on or off, but not the raw audio.
	\item No one (except me) should access raw audio data unless
	it has been transformed from speech to an intent.
	\item The vendor can't switch my light on or off.
\end{enumerate}

\noindent In existing frameworks, developers bear the burden
of protecting user data, and users cannot verify their privacy guarantees:

\paragraph{1) Samsung SmartThings.}
SmartThings~\cite{smartthings}
is an industry home automation framework that can connect hundreds
of brands and thousands of devices. The platform is centered around the
SmartThings Cloud, which remotely manages devices, data, automations, and more.
Devices connect to the Cloud through a phone, hub, WiFi, or third-party cloud.
Vendors write integrations for simple devices such as a light bulb or sensor
based on a schema that also provides authentication and a consistent UI.
Devices with more complex automation logic, such as a smart speaker, must
implement SmartApps, connectors that run on AWS Lambda or another
vendor-hosted server and send events back to the Cloud.

\paragraph{\textit{Vendor Perspective.}}
Amazon Alexa and Google Home are both smart speakers with SmartThings
compatibility. However, they primarily use SmartThings
to proxy commands to hub-connected devices that use different protocols
or schemas. The devices are still connected to their original cloud
services, which they depend on for remote access and tasks like speech-to-text.
While larger companies may appreciate the flexibility of being able to
integrate their
existing cloud infrastructure, smaller vendors face high liability and
startup costs.

\paragraph{\textit{User Perspective.}}
Allowing devices unrestricted network access requires total trust in the vendor.
Network traffic is typically
encrypted, so privacy settings in the companion app are not verifiable.
Raw audio data could be leaked through
vulnerabilities in the smart speaker, SmartApp, Cloud, or the light bulb
the speaker turns on.
The user can restrict communication between different devices through the
SmartThings Cloud, but devices that require additional cloud communication
to provide
functionality are outside the scope of the framework.
None of the four properties can be enforced.

\paragraph{2) Home Assistant.}
Home Assistant (HA)~\cite{homeassistant} is an open-source smart home
framework with a focus on
local control and privacy. The HA architecture depends on an
event bus that fires and listens to events such as state changes and services.
Developers write integrations (e.g., image processing, light, Z-Wave) that
extend the core architecture with small pieces of home automation logic.
Integrations are limited to Python scripts and can access the network.
HA is highly programmable, and users have the option of
self-hosting the framework on local hardware such as a Raspberry Pi.

\paragraph{\textit{Vendor Perspective.}}
HA allows vendors flexibility to execute
functionality locally or in the cloud.
HA offers smart speaker functionality through a combination of multiple
integrations.
Almond~\cite{campagna2017almond} is an open, privacy-preserving virtual
assistant integration that acts as the text-to-intent backend.
Almond also has a repository of apps, called Thingpedia,
such as an app that retrieves the weather.
Other integrations handle speech: Ada~\cite{ada} is powered by Microsoft
Cognitive Services, and Rhasspy~\cite{rhasspy} is an offline
voice assistant.
Thus smaller vendors can easily contribute to some or all parts of a device
simply by writing software integrations.
HA is also compatible with Alexa and Google Home,
which similarly use HA as a hub to access devices like in SmartThings.

\paragraph{\textit{User Perspective.}}
Programmability and the option to self-host make it feasible for HA to
implement the majority of functionality on local hardware,
but in the remaining cases it is hard to verify privacy guarantees.
HA does not sandbox integrations. For example, Ada necessarily
talks to the Microsoft API, but it is unclear if or when other integrations
such as a speech-to-intent integration access the network.
The combination of services, state changes,
and other events make it difficult to manually trace which data might have
been exfiltrated and where. In addition, part of the dataflow for asking
the weather is in a non-ephemeral Almond server rather than an HA integration.
Devices such as the Alexa have similar privacy concerns as in SmartThings.
Thus while the user can take steps to run mostly privacy-preserving software,
it is still difficult to justify every network access that may contain data.
None of the four properties can be enforced.

\paragraph{3) FlowFence.}
FlowFence is a research framework for smart homes that takes a dataflow
approach to privacy instead of access control.
In FlowFence, IoT apps consist of functions that compute on sensitive data
(quarantined modules or QMs), and code that does not compute on sensitive data.
QMs execute inside Java sandboxes on the hub, which can run on an Android phone.
QMs access sensitive data via opaque handles, monitored by the hub
according to flow policies.
Also, QMs communicate via event channels or a key-value store.
Developers declare intended dataflows of sensitive data
to other devices or the network, which users approve via UI prompts
when a flow is first required.

\paragraph{\textit{Vendor Perspective.}}
It is straightforward for vendors to modularize existing
device-side logic and port it to QMs in FlowFence, similar to other
app-based frameworks. However, FlowFence does not explictly
consider how to integrate companion apps into the framework,
such as to remotely turn on a light,
as evidenced by the example flow
that sends door state to the Internet for the user to view.
It also does not consider how to express policies that use cloud services
such as machine learning, which are not
as easily modularized while preserving performance given high
initialization times, and may not fit within the Java programming model.
Vendors must also consider how to set and interpret taint
labels within module code.
Thus FlowFence complicates device-side logic while still requiring
vendors to manage their own infrastructure.

\paragraph{\textit{User Perspective.}}
Unlike the previous frameworks, FlowFence requires all network
connections involving sensitive data to go through the hub.
The user can enforce P1 and P2 by approving flows that
correspond to the properties.
However, it may be difficult for users to understand and approve
every flow, particularly because a device also includes flows for
cloud services and companion apps, which can negate P4.
FlowFence does not have the concept of restricting
categories of flows such as in P3.
Furthermore, some of the suggested flow policies are overly permissive
because they do not include the application semantics in the QM pipeline.
For example, the flow from a speaker to the lock means a malicious
QM could send raw audio to the lock, even if only a subset of that
information is required.\\

\paragraph{4) Karl.} Karl provides privacy through
local hosting that captures \textit{all} the rich interactions that
modern IoT devices have with cloud services, companion apps, and other devices;
and modularity that enables enforceable, fine-grained privacy policies.
Karl's programming model is based off the familiar serverless interface
that is easy for existing vendors to adopt.

\paragraph{\emph{Vendor Perspective.}}
The vendor splits device functionality into firmware for the device,
and modules that cannot run on the device, because they require too
much computation or storage or because they run on a user's phone.
The vendor programs an initial dataflow graph into the firmware,
representing the device's functionality.
Instead of creating a mobile app, the vendor creates a Karl
app, which the hub serves from the user's machine to the user's phone.
The hub downloads modules from a Karl package manager and
executes the modules on local hardware.
The vendor no longer needs to host any cloud infrastructure or user data.

\paragraph{\emph{User Perspective.}}
The user installs a device, which locates the Karl hub on the local network,
and sends the initial dataflow graph.
The user gets a notification that a new device is available and approves it.
As part of approving the device, the user reviews a set of
pipeline permissions corresponding to P1 and P2,
showing which device data may be shared and why.
The user may also assign exit policies such as P3
to device and application data,
preventing sensitive data from being shared under certain conditions
regardless of pipeline permissions.
The user is guaranteed P4 due to local hosting of companion apps,
since the light does not need network access.\\

In the following sections, we describe the programming model for Karl and
illustrate how Karl can enforce novel privacy guarantees that existing
frameworks cannot.

\section{Programming Model}
\label{sec:programming-model}

Karl's programming model is intentionally similar to existing event-driven
programming models to provide a familiar interface to IoT developers.
Unlike existing programming models, Karl's is designed to eliminate
vendor dependence on cloud
infrastructure---cloud services, companion apps, hardware---using
the simplest abstractions for compute and storage.

In this section, we describe our abstractions for compute and storage,
and the Karl app UI.
Karl expresses all functionality as modules that
interact with a persistent data store and Karl web apps.
We then describe the dataflow graphs that
represent device functionality, and discuss example graphs
for three different IoT devices that handle sensitive data.

\paragraph{Serverless modules.}

\begin{table}[t]
    \small
    \begin{center}
    \begin{tabular}{lp{5.8cm}}
    \toprule
    Field & Value\\
    \midrule
    Name & \texttt{light\_switch} \\
    Inputs & \texttt{intent} -- JSON of the form \{ type: ``light", state: <state> \}, where the state is ``on" or ``off"\\
    Outputs & \texttt{state} -- $1$-bit to turn on and $0$-bit to turn off\\
    Domains & N/A \\
    \bottomrule
    \end{tabular}
    \end{center}
    \caption{Example specification of the \texttt{light\_switch} module
    that maps a JSON intent to a state the light bulb can understand.}
    \label{tab:module-specification}
\end{table}

Similar to the serverless programming model, a \textit{module}
is a self-contained program that executes code given some inputs.
Modules can take multiple inputs
and return multiple outputs.
Each input and output is named and associated with
a particular data type, and is part of the module specification
(Table~\ref{tab:module-specification}).
The module must also specify any domain names with which
it requires network access.

Modules execute inside sandboxes managed by the Karl hub.
The module has a single network connection to a sandbox, which
proxies data and network accesses (Listing~\ref{lst:module-sdk})
to the data store and the Internet.
The sandbox decides which accesses are allowed,
forwarding data accesses to a controller
and HTTPS requests to the requested domain.
Note that data accesses are keyed by the names in the module
specification, agnostic of how data is used outside the module.

\begin{lstfloat}[t]
\begin{lstlisting}
class ModuleAPI:
    def read(input, lower_timestamp, upper_timestamp)
    def read_last_n(input, n)
    def read_event(input)
    def push(output, bytes)
    def network(domain, request)
\end{lstlisting}
\captionof{lstlisting}{API used by module code to
read and push to the data store in various ways,
and make HTTPS requests. The API wraps gRPC calls to the sandbox,
which forwards authorized requests.}
\label{lst:module-sdk}
\end{lstfloat}

\begin{lstfloat}[t]
\begin{lstlisting}
import karl
intent = karl.read_event("intent")
if intent["state"] == "on":
    karl.push("state", [1])
else:
    karl.push("state", [0])
\end{lstlisting}
\captionof{lstlisting}{Example implementation of the module in Table~\ref{tab:module-specification}.}
\label{fig:module-spec-implementation}
\end{lstfloat}

\begin{figure*}[t]
    \centering
    \begin{subfigure}[b]{0.25\textwidth}
        \centering
        \includegraphics[width=.85\columnwidth]{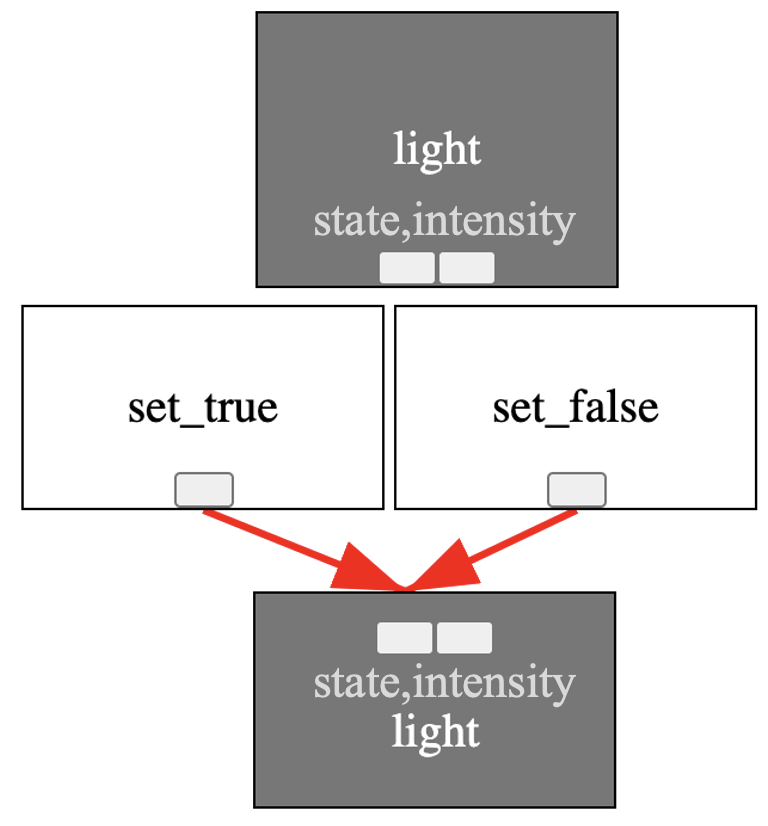}
        \caption{Smart light.}
        \label{fig:dfg-light}
    \end{subfigure}
    \begin{subfigure}[b]{0.26\textwidth}
        \centering
        \includegraphics[width=.85\columnwidth]{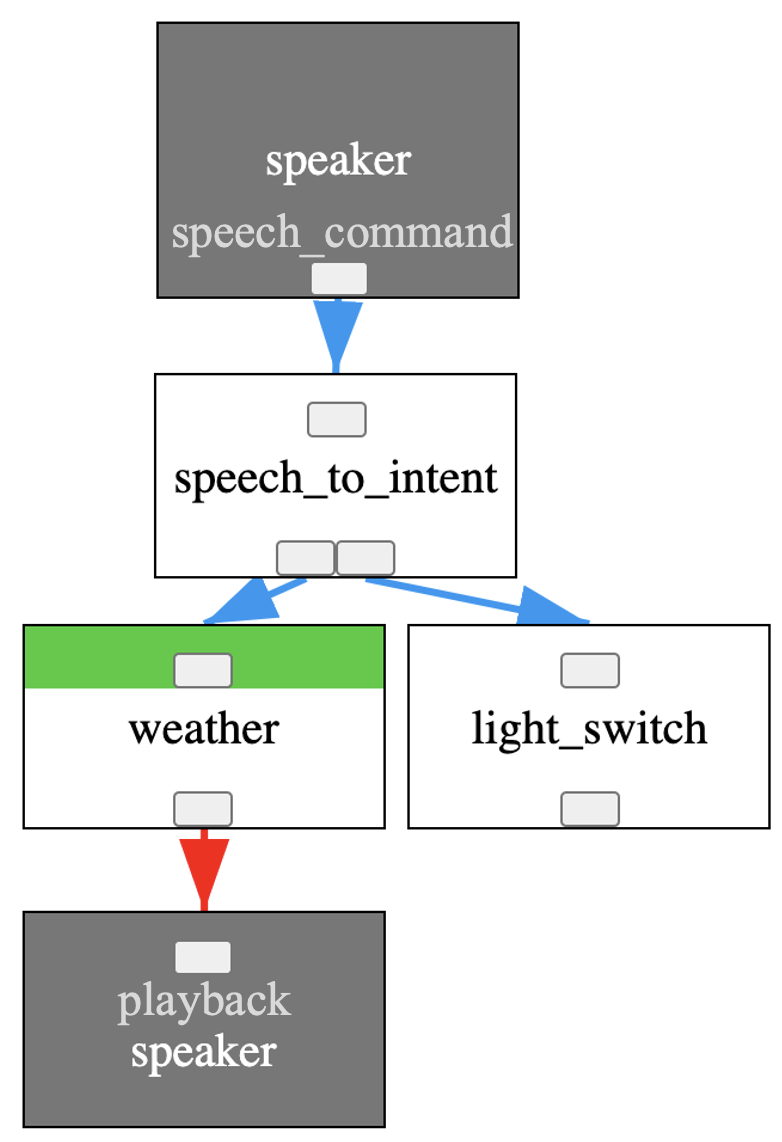}
        \caption{Smart speaker.}
        \label{fig:dfg-speaker}
    \end{subfigure}
    \begin{subfigure}[b]{0.38\textwidth}
        \centering
        \includegraphics[width=.85\columnwidth]{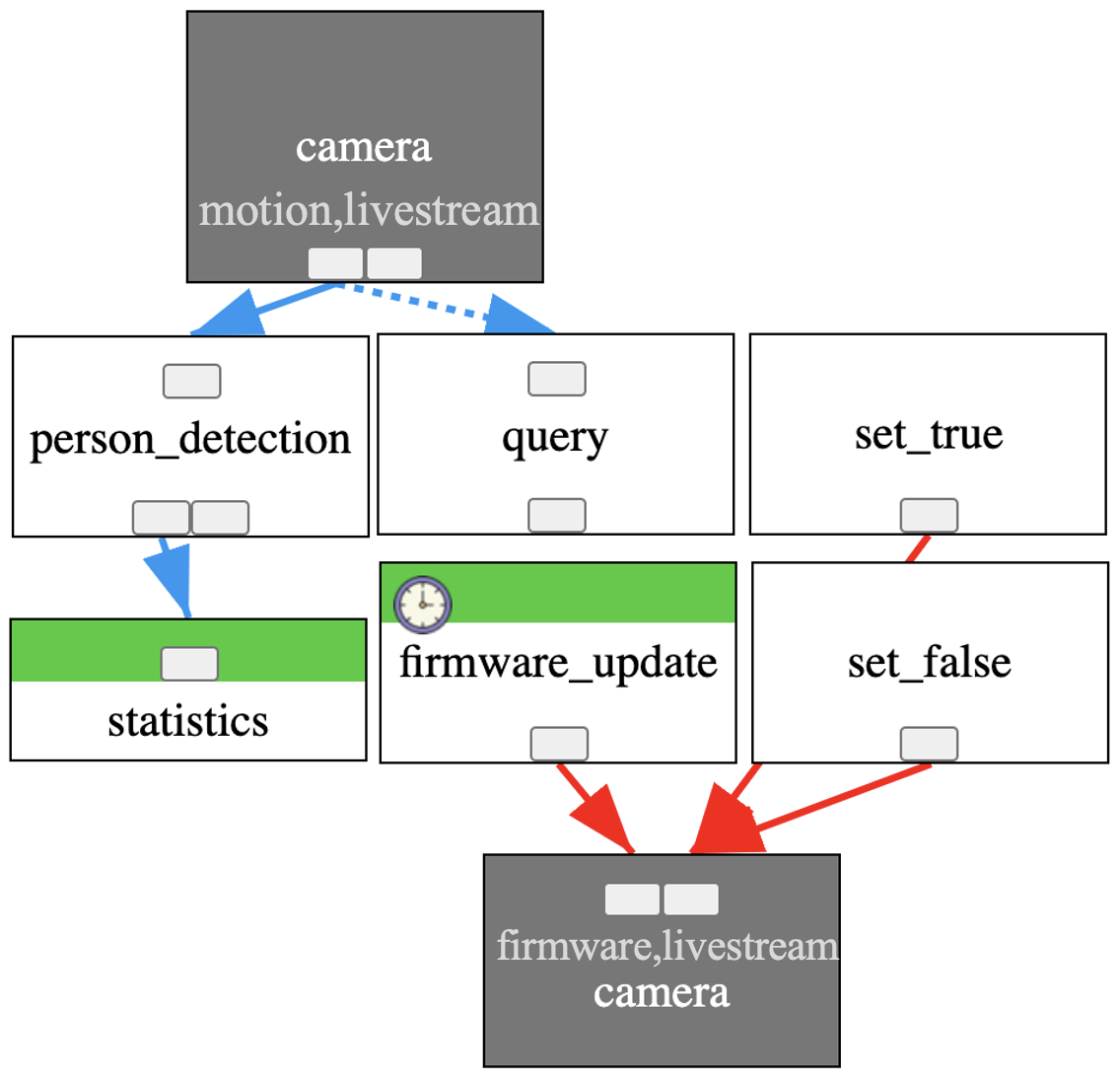}
        \caption{Security camera.}
        \label{fig:dfg-camera}
    \end{subfigure}
    \caption{Example dataflow graphs.
    The gray boxes are devices and white boxes are modules. The smaller boxes
    inside are input and output nodes. Edges connect nodes and
    represent data dependencies. Solid edges are stateless, while dashed
    edges are stateful. Blue edges flow to a module, while red edges flow to a device. Green headers represent modules with network
    access. The clock indicates a fixed interval schedule.}
    \label{fig:dataflow-graphs}
\end{figure*}




\paragraph{Persistent data store.}
The \emph{data store} extends the key-value store, a common
data structure for stateful FaaS~\cite{shillaker2020faasm,sreekanti2020cloudburst}.
In our data store,
keys are \textit{tags} that represent non-hierarchical categories of data
in the global namespace,
and values are append-only logs indexed by timestamps.
While the module API operates in terms of named inputs and outputs,
the data store understands values in terms of tags.
The controller automatically assigns tags to data
when modules read inputs and push outputs
(Listing~\ref{fig:module-spec-implementation}),
based on the dataflow graph (\S\ref{sec:dataflow-graph}).

The purpose of the data store is to provide a simple but
familiar abstraction that allows devices to replicate application logic.
Stateless event-driven programming is
the most common communication pattern
in existing frameworks: SmartThings \texttt{subscribe} API,
Home Assistant event bus, FlowFence event channels.
To provide stateful communication, existing frameworks may use
state or key-value stores associated with the device, or vendor-hosted
cloud storage.

We are able to express a variety of stateful applications using the
data store as the only storage abstraction.
Karl enforces access control rules on tags based on the user's
privacy policies to prevent data
leakage between different modules.
To leverage the privacy properties of stateless applications, which are
ephemeral and cannot leak data between invocations, the data store
additionally limits these modules to reading just the event for a tag
that caused the module to be spawned.

\paragraph{Karl web apps.}
The Karl hub includes a per-user web interface for reading and
pushing to the data store, and configuring and spawning modules.
Vendors can write static pages with JavaScript (i.e. single-page apps)
called \textit{Karl apps}
that wrap these features to replicate the functionality of
mobile companion apps. For example, instead of serving
sensitive photos from vendor-managed cloud storage, the page
can visualize data associated with the corresponding tag in
the data store. Instead of forwarding a request to view a live video
feed through the cloud, the page can spawn a module that turns on the
livestream feature and then visualize data associated with a livestream tag.
It would be a simple extension to also support a stateful
webserver running as a module.

Karl apps allow vendors to provide the equivalent of companion apps
directly through the framework, without hosting cloud infrastructure.
Typically, vendors must host cloud services with longstanding
connections to each device to evade NAT. Alternatively, vendors roll
their own web interfaces on the device. Combined with port forwarding
and insecure authentication, IoT devices can easily expose the home
network to security risks. Thus Karl apps improve privacy and security
compared to the existing state of companion apps.



\subsection{Dataflow Graph}
\label{sec:dataflow-graph}

Each IoT device proposes an initial dataflow graph
where data flows from the device, through modules,
and to other devices and the Internet (Fig.~\ref{fig:dataflow-graphs}).
Though we discuss a visual graph,
devices send their graphs to the hub using JSON.
Karl downloads the requested modules from a Karl
package manager.
Vendors can either upload their own modules or use existing modules
based on their specifications.

The boxes in the dataflow graph are modules and devices,
while nodes are their inputs and outputs.
For example, the light bulb in Fig.~\ref{fig:dfg-light}
outputs its state and intensity, and
has inputs that change the same state.
Each node corresponds to a tag, and
edges represent stateless or stateful data dependencies between nodes.
When the \texttt{set\_true} module pushes to its output
that is connected to a single input,
the controller automatically adds the tag for each node:
\texttt{set\_true.true} and \texttt{\textcolor{purple}{\#light.state}}.
Modules spawn on three different schedules:
when data is pushed to a stateless edge,
at a fixed time interval, or when manually spawned by the user.
Some modules require network access, which implies a flow
to the Internet.

\begin{figure}[t]
    \centering
    \begin{subfigure}[b]{0.48\linewidth}
        \centering
        \includegraphics[width=\linewidth,clip,trim={1cm 1cm 1cm 1cm}]{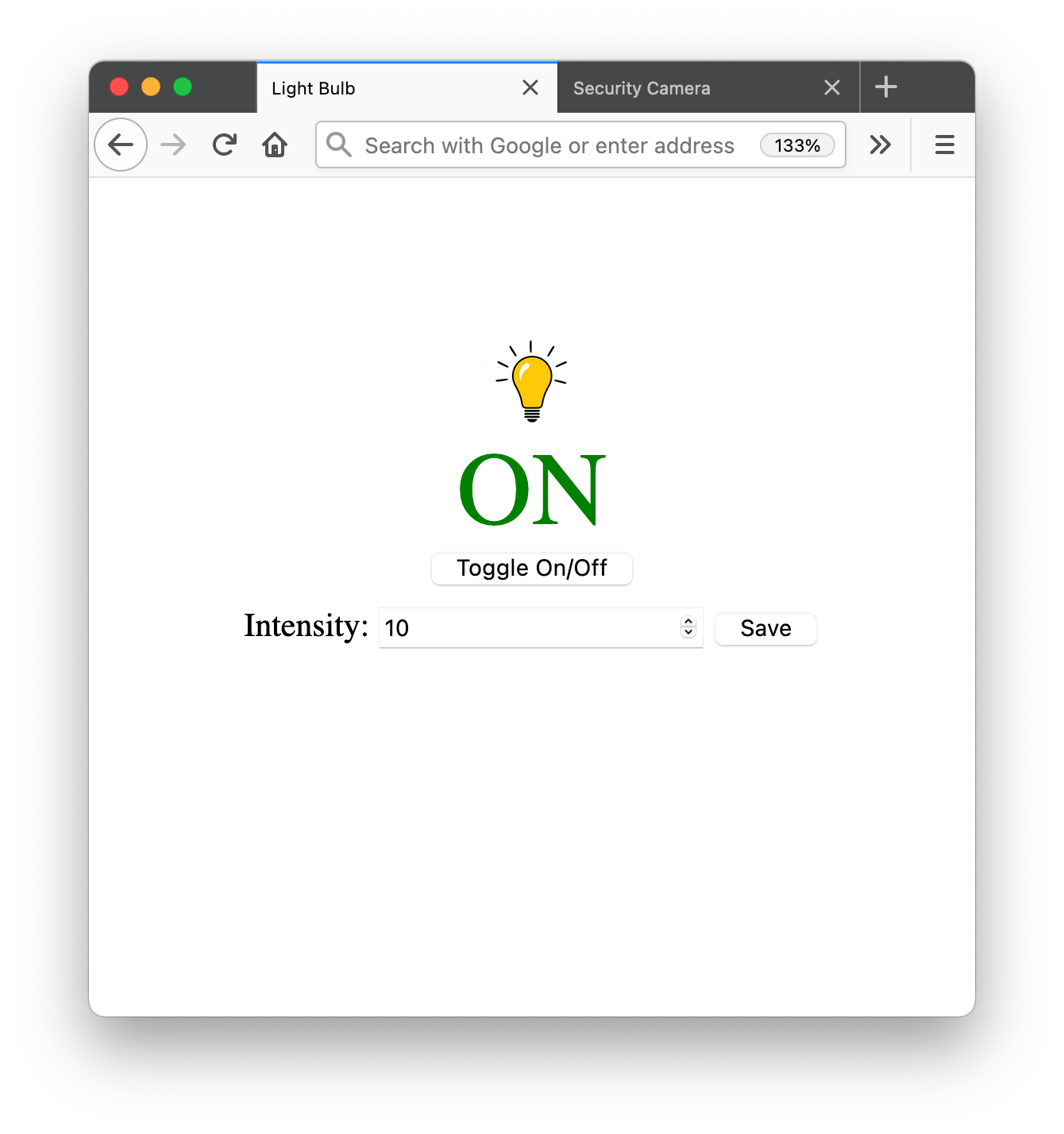}
        \caption{Smart light.}
        \label{fig:app-light}
    \end{subfigure}
    \begin{subfigure}[b]{0.48\linewidth}
        \centering
        \includegraphics[width=\linewidth,clip,trim={1cm 1cm 1cm 1cm}]{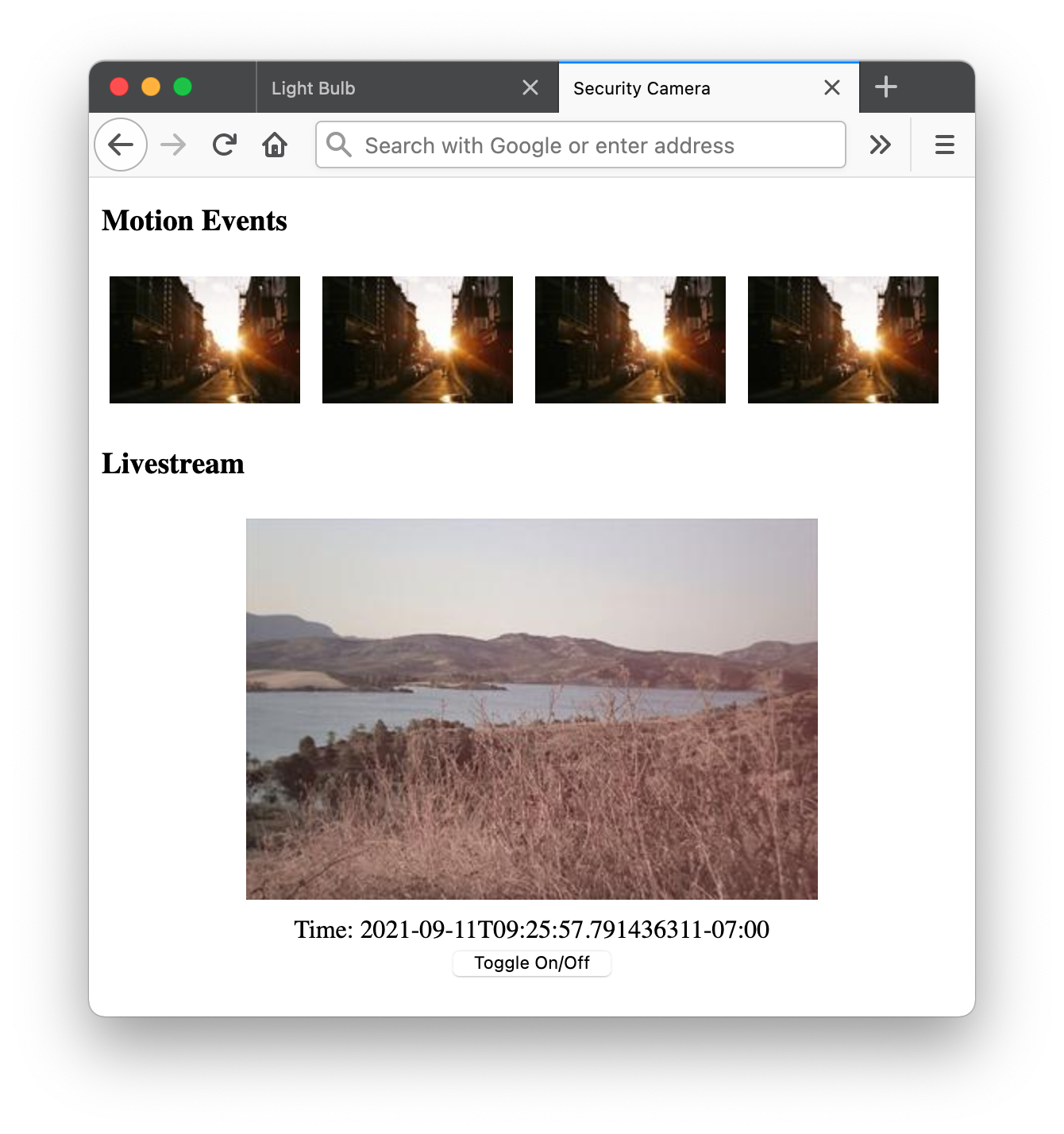}
        \caption{Security camera.}
        \label{fig:app-camera}
    \end{subfigure}
    \caption{Example Karl apps that correspond with the dataflow graphs in Fig.~\ref{fig:dataflow-graphs}.
    The user can set and visualize light state. In the camera, they can see motion detection events and toggle the livestream.}
    \label{fig:karl-apps}
\end{figure}

In addition to data processing flows,
the graph includes the flows needed for Karl apps.
Fig.~\ref{fig:app-light} is the light bulb's Karl app.
The app visualizes the light state and intensity by
reading the sensor output tags. It sets the intensity
by pushing to the \texttt{\textcolor{purple}{\#light.intensity}} input.
It switches the light on or off by spawning a module that pushes a
$1$-bit or $0$-bit to the \texttt{\textcolor{purple}{\#light.on}} input.
These interactions go through user hardware,
so the light bulb state is guaranteed
to be private.

We provide example dataflow graphs for two, more complex devices:
a smart speaker (Fig.~\ref{fig:dfg-speaker})
and a security camera (Fig.~\ref{fig:dfg-camera}).
The smart speaker requires network access to pull weather data,
while the camera pulls firmware updates and pushes analytics.
The smart speaker also expresses
control flow logic using modules.
The camera uses a stateful edge to process a query over
recorded motion events, and spawns a
firmware update module using a fixed interval schedule.

The initial device dataflow graph provides a starting point for users
who introduce new devices to their home.
Users can configure devices to interact with each other,
such as by enabling the smart speaker to turn on the light bulb with
a voice command. The user can also add privacy-preserving modules
between sensitive data and the network to reduce the fidelity of data
before sharing it with the Internet.
We imagine a frontend for privacy settings and automation logic,
such as IFTTT~\cite{ifttt} or the Home Assistant frontend,
that automatically maps user configurations
to the underlying dataflow graph. For example, the frontend could
suggest edges based on corresponding data types or recognize devices
from the same vendor based on overlapping graphs.

\section{Privacy Policies}
\label{sec:privacy-policies}

The modular dataflow graph
provides a convenient foundation for privacy policies based on
fine-grained application semantics.
Karl can regulate the flow of sensitive data from source to sink based on the
semantics of the pipeline of modules in between.
Recall the justifications for IoT network access in Table~\ref{tab:class}.
Karl can provide compute, storage, and connectivity using local
hardware, but devices should still justify when they \textit{must} fundamentally
pull or push data to the Internet.

Karl provides the framework for users to monitor these
dataflows based on two concepts.  \textit{Pipeline permissions} specify
why and to whom data \textit{can} flow based on the pipeline of
modules from source to sink.  They serve to ensure that dataflows
match users' intuitions, and that devices and modules do not engage in
unnecessary communication.  However, seemingly plausible
pipeline permissions can still have unintended consequences.
\textit{Exit policies} capture end-to-end data policies
that block flows unless specific conditions based on modules are met,
regardless of pipeline permissions.


In this section, we discuss how the privacy properties
in Fig.~\ref{fig:privacy-expectations}
map to pipeline permissions and exit policies,
apply these concepts to modify the dataflow graph as a declarative policy,
and describe how Karl compiles the graph
to a non-hierarchical form of mandatory access control.


\begin{figure}[t]
\small
\begin{enumerate}[A.]
\itemsep0em
\item The smart speaker can talk to the weather station if I ask about the weather, without sending the actual recording.
\item The camera can share training data with the vendor.
\item The smart speaker can turn on the light if I tell it to, and without sending the raw audio.
\item The camera can update its firmware from the trusted distribution URL.
\item The camera can share training data with the vendor \textit{only if I am not at home.}
\item The camera can share training data with the vendor \textit{only if it is anonymized} such as by using Prio~\cite{corrigan2017prio}.
\item The light should never need to send data to the Internet.
\item Only I should ever be able to view the livestream.
\item The smart speaker can only send data derived from raw audio if it has first been translated to text.
\item The camera can only send data derived from raw photos if it has been anonymized.
\end{enumerate}
  \caption{Example privacy properties a user might desire.}
  \label{fig:privacy-expectations}
\end{figure}


\subsection{Pipeline Permissions}
\label{sec:pipeline-permissions}

\begin{figure}[t]
\small{
\begin{enumerate}[A.]
	\item\texttt{
		\textcolor{BurntOrange}{speaker.speech\_command}
		$\rightarrow$ speech\_to\_intent
		$\rightarrow$ weather
		$\rightarrow$ \textcolor{ForestGreen}{weather.com}}
	\item\texttt{\textcolor{BurntOrange}{camera.motion}
		$\rightarrow$ person\_detection
		$\rightarrow$ statistics
		$\rightarrow$ \textcolor{ForestGreen}{statistics.com}}
	\item\texttt{\textcolor{BurntOrange}{speaker.speech\_command}
		$\rightarrow$ speech\_to\_intent
		$\rightarrow$ light\_switch
		$\rightarrow$ \textcolor{purple}{\#light.state}}
	\item\texttt{\textcolor{ForestGreen}{firmware.com}
		$\rightarrow$ firmware
		$\rightarrow$ \textcolor{purple}{\#camera.firmware}}
\end{enumerate}

\footnotesize{
Legend=\texttt{\textcolor{BurntOrange}{device.output}},
	\texttt{module},
	\texttt{\textcolor{ForestGreen}{domain\_name}},
	\texttt{\textcolor{purple}{\#device.input}}
	}
	}

\caption{Pipeline permissions that correspond to privacy properties A-D in
Fig.~\ref{fig:privacy-expectations}, detected from the dataflow graphs
in Fig.~\ref{fig:dataflow-graphs}.}
\label{fig:pipeline-permissions}
\end{figure}

\begin{figure}[t]
\small{
\begin{enumerate}[E.]
	\item\texttt{\textcolor{BurntOrange}{camera.motion}
		$\rightarrow$ person\_detection ($+$ \textcolor{BurntOrange}{occupancy\_sensor.at\_home})
		$\rightarrow$ boolean
		$\rightarrow$ statistics
		$\rightarrow$ \textcolor{ForestGreen}{statistics.com}}
\end{enumerate}
\begin{enumerate}[F.]
	\item\texttt{\textcolor{BurntOrange}{camera.motion}
		$\rightarrow$ person\_detection
		$\rightarrow$ prio
		$\rightarrow$ \textcolor{ForestGreen}{a.statistics.com}}\\
	\texttt{$\cdots$
		$\rightarrow$ prio
		$\rightarrow$ \textcolor{ForestGreen}{b.statistics.com}}\\
	\texttt{$\cdots$
		$\rightarrow$ prio
		$\rightarrow$ \textcolor{ForestGreen}{c.statistics.com}}
\end{enumerate}

\footnotesize{
Legend=\texttt{\textcolor{BurntOrange}{device.output}},
	\texttt{module},
	\texttt{\textcolor{ForestGreen}{domain\_name}},
	\texttt{\textcolor{purple}{\#device.input}}
	}
	}

\caption{Pipeline permissions that correspond to privacy properties E and F in
Fig.~\ref{fig:privacy-expectations}. E and F are variants of B if the security
camera had included client-side anonymization in its graph.
}
\label{fig:pipeline-permissions-anonymized}
\end{figure}

When a user registers a device with Karl for the first time,
appending its dataflow graph, they must approve
any new sensitive dataflows as \textit{pipeline permissions}
(Fig.~\ref{fig:pipeline-permissions}).
These permissions ensure there are not any unintended effects
from introducing a new device into a complex ecosystem.
Karl inspects the graph for any new linear paths that exfiltrate
device data to the network (or other devices), or that send data to
the device from the network. Karl presents the textual form of these subgraphs
to users as permissions to review.
These permissions represent tradeoffs between privacy and functionality,
and correspond to intuitive privacy properties.

The user considers the list of pipeline permissions generated by
Karl based on module semantics.
Consider a home with a smart speaker, security camera, and light bulb.
Each device dataflow graph is appended as a disjoint graph.
An automation frontend links \texttt{light\_switch.state}
to \texttt{\textcolor{purple}{\#light.state}}, enabling the user
to turn on the light using a speech command.
The permissions in Fig.~\ref{fig:pipeline-permissions} correspond to
privacy properties A-D in Fig.~\ref{fig:privacy-expectations}.
Permission A explains that the user must send at least
text to the weather station to learn the weather,
while permission D allows a third party to update the camera firmware
because it is a trusted vendor.

The user may reject a pipeline permission if after inspection,
it does not meet their privacy expectations.
For example, given permission B, the user may not think it
necessary to share statistics that could include raw image data.
To gain the user's trust, the vendor might add a
module in the pipeline to make it clear that they use data in
an anonymized form such as E or F
(Fig.~\ref{fig:pipeline-permissions-anonymized}).
There are many such techniques for client-side protection of
data~\cite{jana2013enabling,corrigan2017prio,erlingsson2014rappor,burkhalter2021zeph,ko2018deadbolt}.
In this case, the user may believe it is worth it to share anonymized
data because it improves the product for future use.

Note that the dataflows that access the network need only be
those that fundamentally pull and push data.
This reduces the number of policies the user must consider.
For example, rather than sending the light state to the Internet to
visualize in the companion app, the state goes directly to the Karl app
and the flow does not need to be authorized. Rather than sending raw
audio to a cloud service to process the speech command, Karl translates
the audio locally.

It may be difficult for users to accurately
judge every permission, a common issue in user-driven
policies~\cite{felt2012android,roesner2012user}.
It will be important to conduct user studies to determine how best to present
pipeline permissions in the UI.
But even if the user approves all dataflows,
Karl can still provide auditable logs because it captures every network access
and its provenance.
To provide additional reassurances about sensitive
data, we combine the allow approach of pipeline permissions
with the deny approach using exit policies, in the next section.

\subsection{Exit Policies}
\label{sec:exit-policies}

\begin{table}[ht]
    \centering
    \small
    \begin{tabular}{lll}
        \toprule
         & Tag & Exit Policy\\
        \midrule
        G. & \texttt{\textcolor{BurntOrange}{light.state}} & \textit{false} \\
        H. & \texttt{\textcolor{BurntOrange}{camera.livestream}} & \textit{false} \\
        I. & \texttt{\textcolor{BurntOrange}{speaker.speech\_command}} & \texttt{speech\_to\_intent} \\
        J. & \texttt{person\_detection.image} & $\texttt{boolean}\ |\ \texttt{prio}$ \\
        \bottomrule
    \end{tabular}
    \caption{Exit policies proposed by a user for the devices in Fig.~\ref{fig:dataflow-graphs}, corresponding to privacy properties G-J in Fig.~\ref{fig:privacy-expectations}. The policy defaults to \textit{true} if unspecified, deferring to pipeline permissions.}
    \label{tab:exit-policies}
\end{table}

Even if a user mistakenly approves a pipeline permission,
modularity enables
the user to define \textit{exit policies} on
categories of data (Table~\ref{tab:exit-policies}).
Karl's data store groups data under tags corresponding
to the device it comes from or its downstream
modules. The user then defines conditions under which this data
category can be exfiltrated to the network or another device.
These conditions are at the granularity of Karl's serverless modules,
enabling expressive, high-level policies.

Tags enable data-centric policies on important categories of data,
rather than policies on devices and apps that use the data.
In Karl, devices distinguish the different types of data they push at
the source, such as $\texttt{\textcolor{BurntOrange}{camera.motion}}$
or $\texttt{\textcolor{BurntOrange}{camera.livestream}}$.
Tags also represent the inputs and outputs of downstream modules, such as
$\texttt{indoor\_person\_detection.training\_data}$ for outputs
of a module that runs person detection specifically on indoor cameras.
This distinction between tags is built into Karl's modular programming
model, tightly integrating vendor-defined functionality with policy enforcement.

We provide a simple policy language defined in terms of modules
for specifying conditions under which tags can be exfiltrated.
The language consists of three operators: $\&$, $|$, and $>$.
The $\&$ and $|$ operators represent conjunctions and disjunctions,
while $>$ represents ordering.
Note that $\textit{true}$ implies the exit
policy is met in all conditions and allows all pipelines,
while $\textit{false}$ implies the opposite.
We expect this policy language to be expressive
and intuitive to developers, though a
simpler alternative could use only singular modules.

For example, defining the exit policy
$(\texttt{boolean}\ |\ \texttt{prio}) > \texttt{statistics}$ on the tag
$\texttt{\textcolor{BurntOrange}{camera.motion}}$ indicates
that the camera's motion detection
data can only be exfiltrated in a pipeline that includes one of the
anonymization modules, and then the statistics module.
The \texttt{boolean} module forwards data along
one input if the other input indicates a condition has been met, while
the \texttt{prio}~\cite{corrigan2017prio} module is a technique
for ensuring data is analyzed in aggregate.

Pipeline permissions and exit policies are most useful in
combination. Pipeline permissions enable device functionality
given a privacy tradeoff, while exit policies restrict the
conditions under which data can be exfiltrated.
When two policies conflict, the stricter one is enforced.
For example, Permission B in Fig.~\ref{fig:pipeline-permissions}
conflicts with the exit policy J
in Table~\ref{tab:exit-policies}.
The UI then alerts the user of this conflict to either
remediate the existing pipeline or accept a loss in functionality.

\subsection{Enforcement Mechanism}
\label{sec:enforcement-mechanism}

The primary goal of the enforcement mechanism is simplicity
for the vendors, developers, and users of the framework.
Vendors no longer bear the burden of protecting user data
because the data is locally stored and processed.
Module developers are agnostic of data labels such that they
can write modular functions without considering their global implications.
Karl also uses a simple, non-hierarchical data storage format,
as opposed to accumulating labels as data flows through the graph.
At a high level, Karl creates a modified version of the original
dataflow graph to comply with user-defined policies, then treats
the modified graph as a declarative policy of permitted dataflows,
enforcing access control at the edges.

Karl determines how to modify the dataflow graph based on
pipeline permissions and exit policies.
Karl denies the permissions explicitly denied by the user,
and the ones that conflict with exit policies.
For each denied permission, Karl either revokes network
access or removes the edge to the device input. These changes are
applied in a secondary layer, ensuring the list of permissions presented
to the user remains the same. Denying one permission may affect
an allowed permission, such as if the path overlaps. In this case,
Karl can attempt to duplicate the overlapping
graph such that the policies are independent. If Karl were to
identify the pipeline permissions in the modified graph, they would
include none of the denied permissions and as many of the allowed
permissions as possible, while following exit policies.

We treat the final dataflow graph as a declarative policy of
permitted dataflows between devices, modules, and the Internet.
The resulting graph nicely integrates vendor-defined
functionality with user-defined policies.
Given the final graph, the sandbox enforces access control rules
on the API calls in Listing~\ref{lst:module-sdk}.
When a module pushes data, the sandbox adds tags for the
module's output and the inputs on connected edges.
When a module reads data, it can only read the tags for its inputs.
If the read is along a stateless edge, the module cannot read
data previously pushed to that tag, a property aided by the
ephemerality of modules.
The sandbox allows network accesses only to the domains specified
for that module.

It could be interesting future work to explore how to enforce Karl's policies
using IFC labels instead of non-hierarchical mandatory access control.
Exit policies as Boolean predicates would work nicely
with Boolean labeling schemes~\cite{stefan2011disjunction}.
In this model, one might allow module developers to view the data labels,
though developers could then leak information through the labels.
This design choice could allow greater flexibility in enforcing
policies based on data provenance, at the cost
of complexity~\cite{hunt2018ryoan,fernandes2016flowfence}.
It remains interesting to provide a theoretically-grounded
approach to enforcing high-level policies in a dataflow graph.

\section{Karl Hub}
\label{sec:karl-hub}

\begin{figure}[t]
  \centering
  \includegraphics[width=\linewidth]{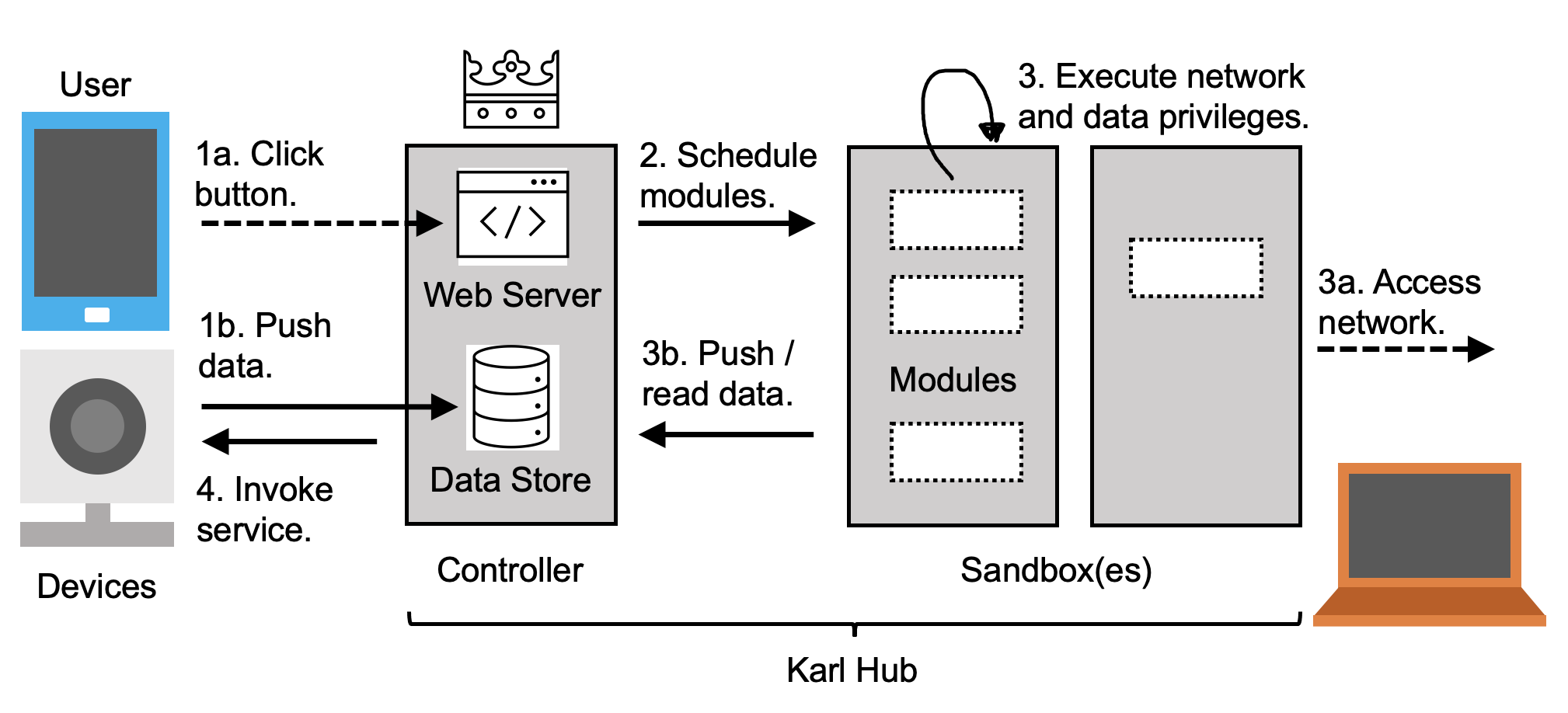}
  \caption{System architecture.}
  \label{fig:karl-hub}
\end{figure}

The disaggregation of storage and compute from the IoT device allows
functionality to exist outside the device in a \textit{Karl hub}.
In contrast, typical frameworks spread the functionality
across local device state, companion apps,
and cloud services, in addition to the hub.

In this section,
we describe how to deploy the hub on local network cloud hardware
managed by the user and discuss the tradeoffs.
We then describe how caching and data
locality optimizations in the scheduler improve the
performance of Karl, particularly for resource-constrained
environments.

\subsection{Deployment}
\label{sec:hub-deployment}

The Karl hub consists of sandboxes that execute Karl modules,
and a controller that schedules computation onto sandboxes
(Fig.~\ref{fig:karl-hub}).
The controller manages the data store, and the webserver
through which users configure and interact with
the smart home using Karl apps.
All data and network accesses from a module must go through
the controller.

\paragraph{Local Deployment.}
The recommended deployment uses local hardware
to best moderate data exfiltration outside the network boundary.
To use the hub, users install a program for the
controller on a local server such as an old laptop.
The program can also come pre-installed on
dedicated hardware.
Then users install sandbox programs on the same server (or others),
making up the computational capacity of the hub.

\paragraph{Cloud Deployment.}
If resources are limited and scalability is an issue,
it is possible to host the hub on cloud hardware that is still
managed by the user.
The user can rent a cloud server that meets the smart home's
computational demands at comparable cost to
IoT cloud subscriptions, with better privacy.
In future work, we hope to provide controller drivers for any
backend that implements the data store and module abstractions.
This can include serverless platforms like AWS Lambda or even
community Karl sandboxes. Then the smart home can primarily leverage
local hardware but rollover to other hardware in the event of
resource constraints.

\subsection{Scheduler}
\label{sec:scheduler}

To improve performance in resource-constrained
environments, we provide a scheduler that optimizes for the smart home.
In particular, writing stateful cloud services with large
data dependencies, such as person detection and other ML tasks,
as serverless modules
results in high initialization and data transfer times.
In response, we implement data caching optimizations and
modify the scheduler to consider data locality to reduce
the impact of these modules on end user latency.

\paragraph{Cold-cache.}
The \textit{cold-cache} optimization mitigates the effect of large
data dependencies on increased network latencies.
ML modules may contain large models, particularly
if they were offloaded to obviate IoT space constraints.
Modules written in interpreted languages may also contain large
code dependencies.
The sandbox unpacks the files in a local filesystem cache, and mounts repeated
modules using \textit{aufs}, an overlay filesystem, to not modify the root.
To avoid sending repeat dependencies, the controller tracks which modules it
has spawned on which sandboxes.
If the sandbox has evicted the module, the controller
takes note and tries again.

\paragraph{Warm-cache.}
The \textit{warm-cache} optimization minimizes long intialization times
by preemptively launching slow modules and pausing until the first
data or network access. The sandbox then waits for the controller
to spawn the same module and continues executing at this point.
An ML module might preemptively load its model in the initialization
phase, and handle an inference request as soon as the module is re-spawned.
Modules cannot be long-running services themselves, as they would be
unnecessarily stateful and leak data across invocations.\\

The scheduler manages a queue of modules that are ready to be
spawned and decides which sandbox to send each module to.
Modules enter the queue based on the schedule assigned to the module:
when data is pushed to a particular tag, at
a regular interval (e.g., daily), or when manually triggered
through the webserver.
Given a module, the scheduler selects a sandbox without an active request,
prioritizing those with the module cached.
If a request does not return after a certain
timeout, the scheduler cancels the request and retries on a different
sandbox with an exponentially increasing timeout.

Though our scheduler prioritizes caching
and availability, the scheduler could consider many other
factors to best utilize the resources available to it. One example
is to collect statistics about which sandboxes are fastest for specific
modules, and map modules to that sandbox. This is particularly important
on heterogeneous hardware. If Karl used a combination of local and cloud
hardware, the scheduler could account for privacy concerns on cloud
hardware,
and data locality for latency. The scheduler could balance modules
based on whether the request needs real-time latency for a user
interaction or better hardware for a computationally expensive task.






\section{Implementation}

\begin{table*}[t]
\small
\centering
  \begin{tabular}{lllllll}
    \toprule
    Module Name&LOC&Size&Language&Inputs&Outputs&Domains\\
    \midrule
    boolean           & 27 & 2.0MB & Rust & condition,input & output & -\\
    firmware\_update  & 11 & 2.0MB & Rust & - & firmware & firmware.com\\
    light\_switch     & 24 & 2.0MB & Rust & light\_intent & state & -\\
    person\_detection & 46 & 981MB & Python & image & training\_data,count & -\\
    speech\_to\_intent & 75 & 38MB & Python & speech & weather\_intent,light\_intent & -\\
    query             & 36 & 2.1MB & Rust & image\_data & result & -\\
    set\_true         & 5  & 2.0MB & Rust & - & true & -\\
    set\_false        & 5  & 2.0MB & Rust & - & false & -\\
    statistics        & 12 & 2.0MB & Rust & data & - & statistics.com\\
    weather           & 25 & 2.1MB & Rust & weather\_intent & weather & weather.com\\
  \bottomrule
\end{tabular}
  \caption{Module implementations and specifications.
  The Rust modules are $\approx$2.0MB because
  they all statically compile MUSL libc. The Python modules are larger
  because Python is an interpreted language with more dependencies.}
  \label{tab:module-implementation}
\end{table*}

We build a prototype of Karl in Rust, with SDKs in Rust and Python, in $\sim$7000
LOC\@.
The controller, sandboxes, devices, and modules communicate over gRPC.
The webserver passes tags and module names to Karl web apps via
the Handlebars templating language, authenticates each app with cookies,
and isolates app data from the Internet using Content Security Policy.
Module sandboxes use Firejail,
an SUID program based on jails and namespaces, and a Netfilter firewall.

We implemented 4 devices---a smart speaker, a light, a camera and
an occupancy sensor---and 10 modules ranging from ML
to multi-device interactions, shown in
Table~\ref{tab:module-implementation}.
We use Mask R-CNN~\cite{he2017maskrcnn} for person detection and
Picovoice~\cite{picovoice} for speech-to-intent.
We modeled devices as programs that push data to the hub at a
fixed interval. The camera pushes 156KB PNG images, and the microphone pushes
172KB WAV audio files.
We were able to put Karl camera firmware on a hacked WyzeCam
v2~\cite{xiaomi-dafang},
and a Karl smart speaker on a Raspberry Pi v4 with hardware accessories.

\begin{figure}[t]
\centering
  \begin{subfigure}[b]{0.42\columnwidth}
    \centering
    \includegraphics[width=\linewidth]{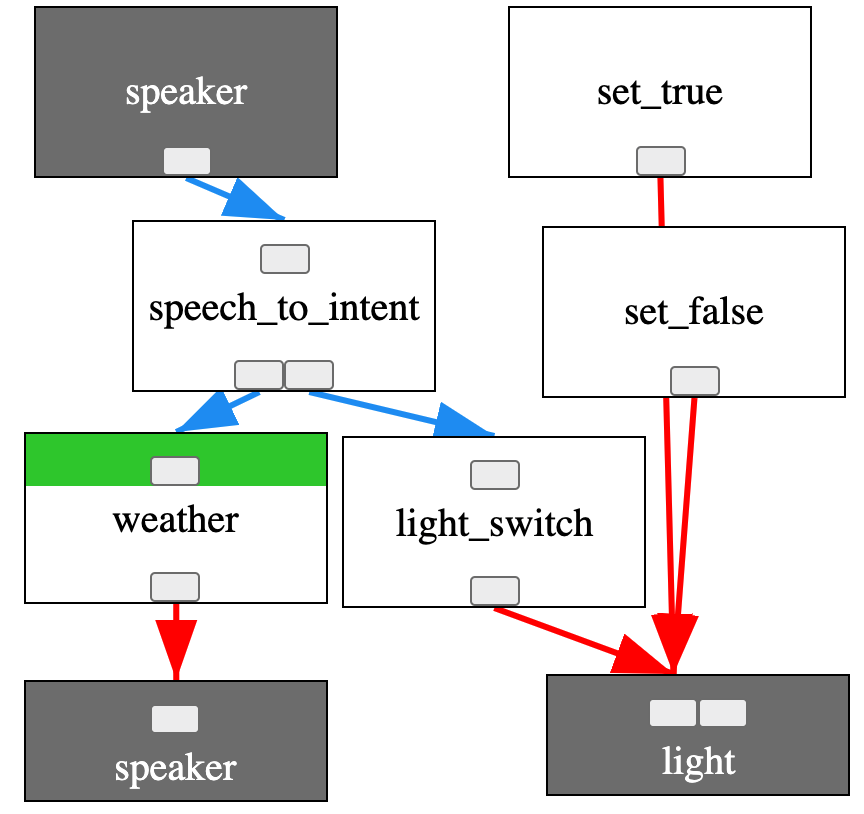}
    \caption{}
  \end{subfigure}
  \begin{subfigure}[b]{0.55\columnwidth}
    \centering
    \includegraphics[width=\linewidth]{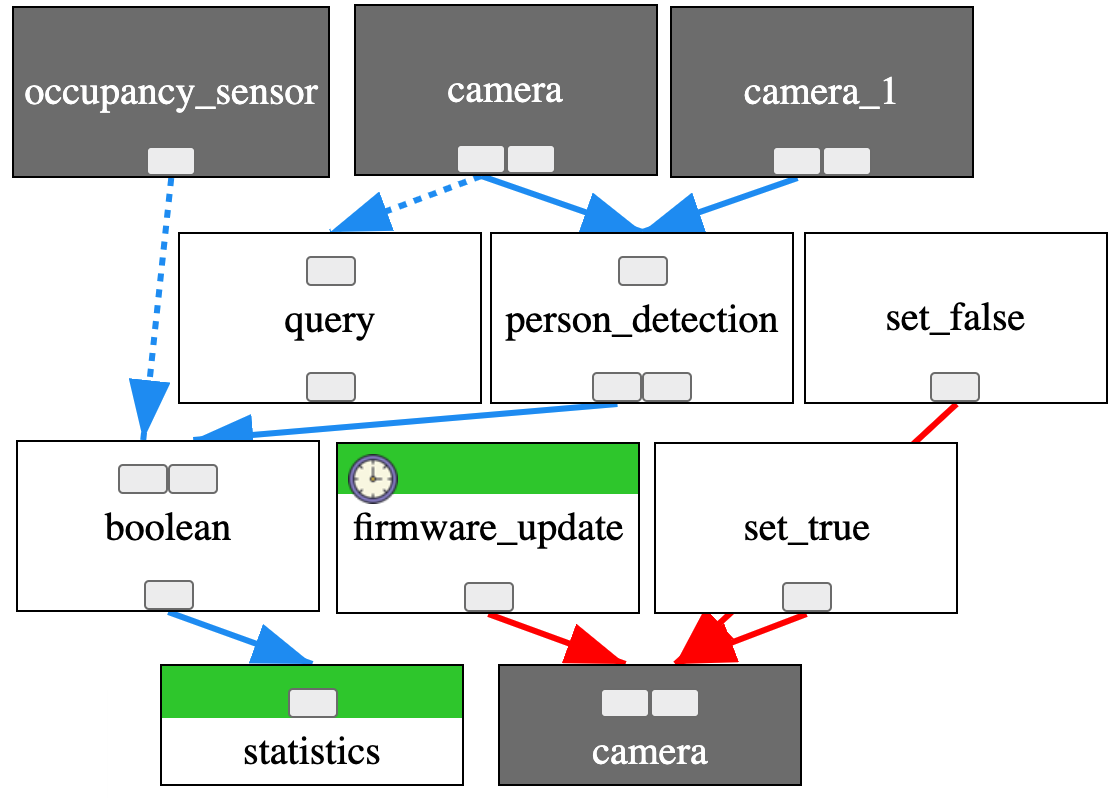}
    \caption{}
  \end{subfigure}
  \caption{End-to-end applications with a speaker and a light (left),
  and an occupancy sensor and two cameras (right).}
\label{fig:experimental-setup}
\end{figure}

We combined these devices and modules in two end-to-end applications
to highlight the range of functionality that Karl can express
(Fig.~\ref{fig:experimental-setup}).
The first application supports speech commands to a smart speaker that
looks up the weather or turns on a light.
The second runs person detection when motion is detected from the camera,
sends training data to the Internet only when the user is not home,
and allows the camera to check for firmware updates.



\section{Evaluation}
\label{sec:evaluation}

We evaluate Karl to answer four questions:

\begin{itemize}
\itemsep 0pt
  \item Can Karl preserve existing device functionality in device-side logic, mobile companion apps, and cloud services?
  \item Can Karl enforce useful privacy policies that are difficult to enforce in existing frameworks?
  \item Does Karl provide reasonable latency for real-time interactions and ML applications on local hardware?
  \item How does hosting Karl on cloud hardware affect performance and cost?
\end{itemize}


\iffalse

\noindent We ran our controller on Ubuntu 20.04 on a 10-year-old
machine with two 4-core Xeon E5620 CPUs and 48GiB DDR3 RAM, typical of
discarded last-generation servers.  We ran a single sandbox and
emulated devices on the same server.

\else

\noindent We run the controller under Ubuntu 20.04 on a 10-year-old
machine with two 4-core Xeon E5620 CPUs typical of discarded
last-generation servers (CPUs \$8/pair on eBay) and 48GiB of RAM
(mostly unused, as seen in Table~\ref{tab:cloud-metrics}).  We run a
single sandbox and emulate the devices on the same server.

\fi

\subsection{Does Karl's serverless programming model preserve device functionality?}

The applications we implemented in Fig.~\ref{fig:experimental-setup}
demonstrate that Karl can preserve the functionality of
existing IoT devices in its programming model.
We
are able to express device-side logic, companion apps, and cloud services
using local hardware under a single framework, unlike existing systems
where IoT devices are still tied   to vendor infrastructure.

We found device-side logic to be modular and event-driven,
fitting with Karl's modules and stateless edges.
We designed edges to use common data types such as image
formats or JSON, such that even modules from different
vendors could be compatible. Otherwise,
one could implement connector modules that convert between formats.
We used the \texttt{boolean} module to express control flow logic
such as conditional policies on whether the home was occupied or
the time of day.

We designed Karl apps to match existing companion apps
without a cloud mediator.
The light app lets the user adjust the light
with low latency because it communicates on the
local network.
The camera app lets the user view a livestream on the user's phone,
preventing breaches where employees or hackers of a vendor service see
the feed~\cite{ringsnooping2020,verkada2021}.
Storage is also controlled by the user and kept on local hardware.

We implemented two ML modules that typically run as stateful cloud
services: speech-to-intent and person detection.
When searching for ML implementations, we were pleasantly surprised by the variety of
open-source options~\cite{picovoice,rhasspy,he2017maskrcnn,hannun2014deep},
depending on how users prioritize latency or accuracy.
Though ML typically runs on dedicated CPU and GPU servers in the cloud, our experiments showed
reasonable latency on our hub's 10-year-old CPU.
In general, we believe that ML applications previously thought to be too costly for
local hardware will be cheaper to support in the future, as commodity hardware vendors are
implementing ML acceleration, e.g. in Intel's integrated GPUs and Apple's Neural Engine.

\subsection{Can Karl enforce useful privacy policies?}

To determine what is considered ``useful", we analyzed Karl in terms of
existing studies on smart home users.
Zheng et al.~\cite{zheng2018user} suggests that users care strongly
about audio-visual data, would prefer sharing data in aggregate,
and prioritize functionality over privacy.
Dixon et al.~\cite{dixon2012operating} recommends time-based access
control and extra sensitive devices as security primitives, and a
central management layer.
The Karl hub can enforce these properties with exit policies on tags for
audio-visual or sensitive data, conditions based on anonymization
or boolean modules for time-based policies,
and a framework that that tightly integrates functionality and privacy.

We validated Karl's generality by implementing all the privacy properties in Fig.~\ref{fig:privacy-expectations}.
Existing research frameworks either cannot express or enforce all these policies (\S\ref{sec:overview}).
Hubs integrate local devices, but do not account for data
handled in companion apps and vendor
services~\cite{wang2018fear,fernandes2016flowfence,tian2017smartauth}.
Another reason is the granularity at which frameworks define policies.
Low-level IPCs lack semantic meaning~\cite{jia2017contexiot},
while high-level app descriptions do not directly match
functionality~\cite{tian2017smartauth}.

Next, we discuss the enforcement mechanism and how it
reflects the users' privacy expectations.
In many cases, denying a pipeline permission does not affect
other pipelines, such as when we denied the speaker from turning on the
light bulb in Fig.~\ref{fig:experimental-setup}. In other cases,
it affects other permissions that the user has allowed,
leading to unintended side effects. When we denied the occupancy sensor
from sending data to statistics.com, Karl revoked network access
from the \texttt{statistics} module, and the camera was no longer able
to share training data. This side effect makes sense---when
we decided not to leak metadata about our occupancy, the \texttt{boolean}
module could no longer properly anonymize training data, so the camera pipeline
was also denied. The UI indicates when pipeline permissions
conflict with exit policies and each other to convey side effects to
users, though user studies should be performed to improve the design.
Karl still provides the foundation on which IoT hubs can build
to express and enforce comprehensive privacy policies on
modular application semantics.

\subsection{How is performance on local hardware?}

\begin{figure}[t]
  \centering
  \includegraphics[width=0.4\linewidth,clip,trim={0 10cm 0 0}]{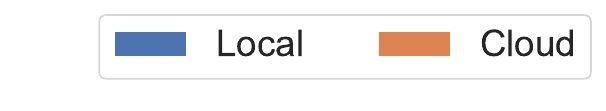} \\
  \includegraphics[width=0.31\linewidth]{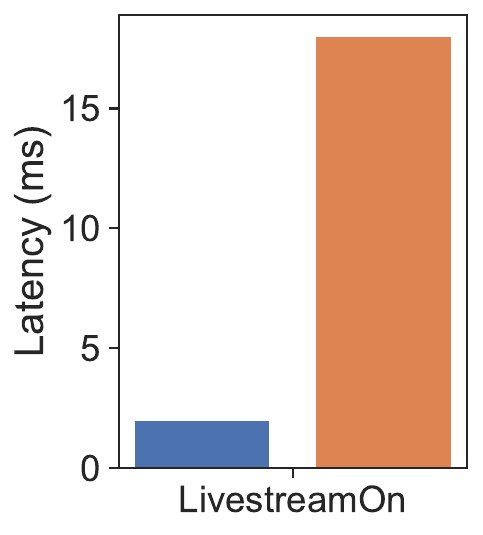}
  \includegraphics[width=0.32\linewidth]{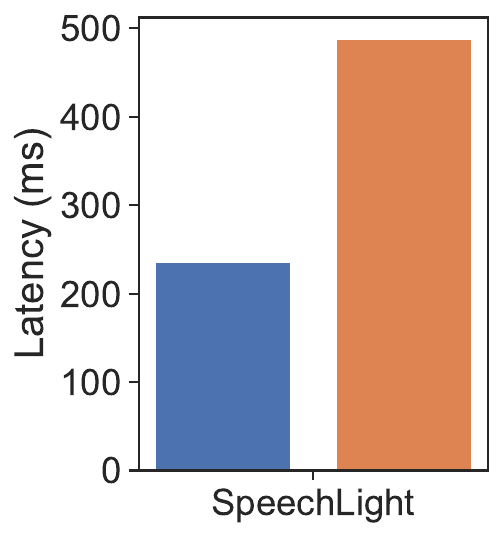}
  \includegraphics[width=0.30\linewidth]{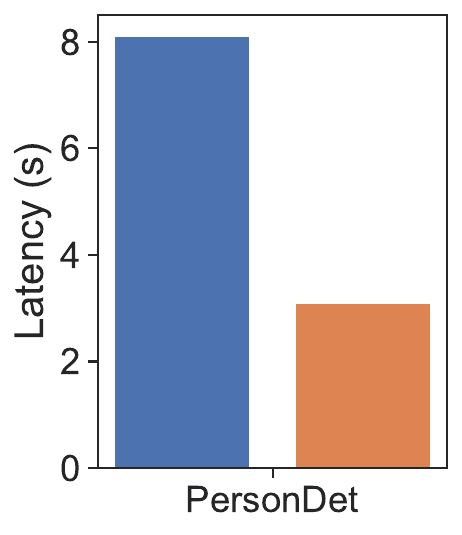}
  \caption{End-to-end latencies on local vs. cloud deployment
  with warm-cache optimization. AVX2 support on cloud hardware
  offsets WAN latencies in PersonDet (Pipeline III\@).}
  \label{fig:local-vs-cloud}
\end{figure}

\begin{table*}[t]
  \centering
  \small
  \begin{tabular}{lp{12cm}ll}
    \toprule
     & Pipeline & Real-time? & ML?\\
    \midrule
    LivestreamOn (I) & \texttt{set\_true $\rightarrow$ \textcolor{purple}{camera.livestream}} & yes & no\\
    SpeechLight (II) & \texttt{\textcolor{BurntOrange}{speaker.speech\_command} $\rightarrow$ speech\_to\_intent $\rightarrow$ light\_switch $\rightarrow$ \textcolor{purple}{light.state}} & yes & yes\\
    PersonDet (III) & \texttt{\textcolor{BurntOrange}{camera.motion} $\rightarrow$ person\_detection ($+$ \textcolor{BurntOrange}{occupancy\_sensor.at\_home}) $\rightarrow$ boolean $\rightarrow$ statistics $\rightarrow$ \textcolor{ForestGreen}{statistics.com}} & no & yes\\
    \bottomrule
  \end{tabular}
  \caption{Evaluated pipelines for end-to-end latency.}
  \label{tab:pp-evaluation}
\end{table*}

Fig.~\ref{fig:local-vs-cloud} demonstrates that Karl has
reasonable end-to-end latency
for real-time user interaction and computational tasks in a
resource-constrained environment. We evaluated combinations of these
two types of requests using the pipelines in Table~\ref{tab:pp-evaluation}.
We measured end-to-end latency as the time
from when a user interacts with system to when the user can observe its
intended effect. Latencies are the average of 5 trials, after warmup.
All performance optimizations are enabled.

We tested two real-time user interactions and observed they are within a
reasonable human response time.
We turned on a camera livestream through a Karl app in 2ms,
and turned on a light through a speech command in 236ms.
In comparison, we asked an Amazon Alexa the weather as a simple experiment
and only received a response $2.1$s after finishing the command.
Device usability relies on real-time interactions, an area where
Karl's local computation excels.

Computational tasks can be constrained by user hardware,
though they do not necessarily require real-time latency.
Person detection and post-processing took $8.1$s,
bottlenecked by the lack of AVX2 instructions in our hardware
(it took $3.1$s on cloud hardware with AVX2).
Unless the Karl hub is fully saturated, we do not
expect occasionally compute-heavy tasks like this to significantly
impact user experience.



\begin{table}[t]
  \centering
  \small
  \begin{subfigure}[b]{\linewidth}
    \begin{tabular}{llll}
      \toprule
      & Baseline & Cold-cache & Warm-cache \\
      \midrule
      LivestreamOn & 64ms & 54ms (\textcolor{ForestGreen}{16\%}) & 2ms (\textcolor{ForestGreen}{97\%}) \\
      SpeechLight  & 740ms & 515ms (\textcolor{ForestGreen}{30\%}) & 236ms (\textcolor{ForestGreen}{68\%}) \\
      PersonDet    & 17.0s & 11.1s (\textcolor{ForestGreen}{34\%}) & 8.1s (\textcolor{ForestGreen}{53\%}) \\
      \bottomrule
    \end{tabular}
    \caption{Local deployment.}
    \label{tab:latencies-local}
  \end{subfigure}
  \begin{subfigure}[b]{\linewidth}
    \begin{tabular}{llll}
      \toprule
      & Baseline & Cold-cache & Warm-cache \\
      \midrule
LivestreamOn & 126ms & 99ms (\textcolor{ForestGreen}{21\%}) & 18ms (\textcolor{ForestGreen}{86\%}) \\
SpeechLight & 1110ms & 894ms (\textcolor{ForestGreen}{19\%}) & 488ms (\textcolor{ForestGreen}{56\%}) \\
PersonDet & 10.1s & 5.2s (\textcolor{ForestGreen}{49\%}) & 3.1s (\textcolor{ForestGreen}{69\%}) \\
      \bottomrule
    \end{tabular}
    \caption{Cloud deployment.}
    \label{tab:latencies-cloud}
  \end{subfigure}
  \caption{Effect of caching optimizations on end-to-end latencies of
  the three pipelines in Table~\ref{tab:pp-evaluation},
  where the baseline is Karl with no optimizations.
  In parentheses, percent improvement over baseline.}
  \label{tab:opt-latencies}
\end{table}


Table~\ref{tab:latencies-local} demonstrates how our caching optimizations
improved performance.
The cold-cache optimization most benefited Pipeline III
by eliminating the time
to resend ML data dependencies and Python libraries over the network.
The size of the \texttt{person\_detection} module was
$490\times$ the size of the \texttt{set\_true} module in Pipeline I,
which benefited the least.
The warm-cache optimization most benefited Pipeline I, which had
long initialization times for mounting the filesystem and establishing
a sandbox, relative to the remaining execution. In Pipeline III,
warm-cache eliminated the pre-processing
time of loading a model, reducing latency by $3.0$s.


\subsection{How does hosting Karl in the cloud affect performance and cost?}

\begin{table}[t]
  \centering
  \small
  \begin{tabular}{llll}
    \toprule
    & Controller &  Host &    Host\\
    & (Max Mem.) &  (Max Mem.) & CPU\%\\
    \midrule
    Baseline   & 8.7 MB  & 8.6 MB   &     0\% \\
    I (warm)   & 14.7 MB & 15.3 MB   &    3\% \\
    I (cold)   & 14.5 MB & 13.0 MB   &    3\% \\
    II (warm)  & 1.25 GB & 938 MB   &     9\% \\
    II (cold)  & 1.26 GB & 846 MB   &     9\% \\
    III (warm) & 3.03 GB & 2.04 GB   &    92\% \\
    III (cold) & 3.03 GB & 2.02 GB   &    92\% \\
    \bottomrule
  \end{tabular}
  \caption{Resource utilization metrics for each pipeline running
  alone on cloud hardware. The baseline is measured before
  registering any devices. Most memory
usage comes from writing data dependencies from the network request to disk
in the initial cold cache request. The warm cache keeps active modules in memory.}
  \label{tab:cloud-metrics}
\end{table}

We compared the network latency of the same pipelines using Karl deployed
on user-managed cloud hardware. We used an m510 CloudLab server
with an Intel Xeon D-1548 processor with 8 cores and 64GiB RAM\@.
We observed 16ms more latency from Pipeline I,
which sent a request to the camera over the WAN with larger and more
variable latencies.
Module data accesses still went over the same LAN as the hub.

Another aspect of performance is the hardware. Notably, the processor
on the CloudLab server supports AVX2. If the user has
particular applications they want to run that rely on special hardware for
performance, cloud platforms could give them flexibility in selecting
exactly the resources they need.

A back-of-the-envelope calculation finds that we can deploy Karl
in AWS for \$14.80/month, which is comparable to IoT subscriptions
like Ring Protect for \$10/month with greater privacy.
Most IoT traffic is frequent and low-bandwidth~\cite{oconnor2019homesnitch}
similar to Pipeline I, so we assume the bottlenecks to involve ML\@.
Given the CPU and memory usage statistics of each
pipeline (Table~\ref{tab:cloud-metrics}),
we select an instance that can handle Pipeline III\@.
Even if multiple cameras handle 100 requests/day,
these tasks are hardly enough to fully utilize a sandbox if they rarely overlap.
On AWS, a t2.medium reserved instance with 2 vCPUs and 4.0 GiB RAM costs
\$14/month, and 10 GB of EBS storage is \$0.80/month.
We hope to reduce these costs in future work by leveraging
\textit{serverless} cloud platforms for Karl sandboxes instead~\cite{palkar2017diy}.

\label{sec:discussion}
\section{Discussion}

In this section, we discuss some of the more practical
concerns about adopting Karl.

\paragraph{Administrative benefits of local hosting.}
We argue the convenience of cloud hosting is overstated.
The Karl hub has the same uptime as the home router on which it
depends for connectivity, and is not affected by cloud
outages~\cite{awsoutage2020,awsoutage2021,daoudi2020summer,zdnet2021osborne}
Another issue Karl addresses is the heterogeneity of
individual vendors~\cite{ubiquiti2021krebs,garlati2016owlet,larson2017fda},
who can focus on building software and hardware
rather than rebuilding the same infrastructure. Karl can integrate
security best practices such as multi-factor authentication and
encrypted network protocols directly into the framework, and durability
measures such as encrypted backups on cloud or decentralized
storage~\cite{mashtizadeh2013replication,benet2014ipfs}.

\paragraph{Vendor incentives to adopt Karl.}
Vendors can provide users lower latency
for interactive applications and real-time processing in
bandwidth-limited wide-area networks, similar to
edge computing~\cite{satyanarayanan2017emergence}.
Smaller vendors in particular would benefit from lower upfront costs
without having to recoup the costs of hosting cloud infrastructure,
and not having to invest in complying with stricter legislation trends.

\paragraph{Business model other than data monetization.}
We imagine a world in which users are willing to pay a privacy premium
for just hardware devices and software modules, without sacrificing
their data. It is reasonable to run proprietary software
on user hardware, as in the mobile app market.
Karl also does not prevent vendors from collecting data,
as long as they do it transparently and with justification. Karl is
particularly compatible with client-side anonymization
techniques~\cite{jana2013enabling,corrigan2017prio,erlingsson2014rappor,burkhalter2021zeph,ko2018deadbolt}.


\paragraph{Privacy legislation trends.}
There has been a trend towards stricter privacy laws such
as GDPR~\cite{gdpr,ccpa}.
Karl automatically ensures privacy through local hosting
so vendors without the security expertise to comply
can focus on building software and hardware.
In general, privacy laws need to strike a balance between what
protects user data most and what is practical for vendors to implement.
We hope Karl and future research can influence legislation by
demonstrating the practicality of more privacy-friendly approaches.

\paragraph{Transition strategy.}
The easiest way to transition to Karl would be for existing frameworks
such as SmartThings and Home Assistant to adopt its ideas.
Karl's event-driven programming model shares many similarities, but
the difference is how these frameworks leverage local hosting and modularity
to provide privacy guarantees.
Existing frameworks should be more restrictive about their concept of
a ``module'', and its data and network privileges.
They should determine all pathways for sensitive data to be leaked,
including the companion apps, cloud services, and multi-device
interactions of modern devices.
Adopting these ideas can provide the foundation for privacy policies
such as pipeline permissions and exit policies that can be enforced
and expressed by the user.

\section{Conclusion}

IoT devices that belong to end users should not depend on
vendor-maintained infrastructure to operate.
Karl provides a better model that prioritizes privacy while
preserving the functionality of modern devices:  IoT devices
outsource computation and storage to user-controlled hardware,
on which Karl sandboxes software modules from different vendors.
With Karl, vendors extricate themselves from hosting cloud infrastructure,
and users can
unilaterally enforce coherent security across devices and modules.
Karl introduces two security mechanisms:
\emph{pipeline permissions}
that permit device data to be shared given some justification
and \emph{exit policies} that block flows unless specific conditions are
met.
We demonstrate Karl's viability through several IoT applications
with comparable performance and much greater privacy.

\section*{Acknowledgments}
This research was supported in part by affiliate members and other supporters of
the Stanford DAWN project (Google, VMWare, Ant Financial and Meta), and by the
NSF under Grant No. 1931750 and 1900638 and CAREER grant CNS-1651570. Any
opinions, findings, and conclusions or recommendations expressed in this
material are those of the author(s) and do not necessarily reflect the views of
the National Science Foundation.

\bibliographystyle{plain}
\bibliography{karl}

\end{document}